\documentclass[aps,preprintnumbers,twocolumn,amsmath,amssymb,prb,showpacs]{revtex4}
\usepackage{bbm}
\usepackage{mathrsfs}
\usepackage{amsmath}
\usepackage{graphicx}
\usepackage{ulem}
\usepackage{color}
\newcommand{\be}{\begin{equation}}
\newcommand{\ee}{\end{equation}}
\newcommand{\bea}{\begin{eqnarray}}
\newcommand{\eea}{\end{eqnarray}}
\newcommand{\bes}{\begin{split}}
\newcommand{\ees}{\end{split}}

\renewcommand{\vec}[1]{\mathbf{#1}}

\newcommand{\tr}{\operatorname{Tr}}

\newcommand{\bs}{\boldsymbol}

\begin{document}
	\title{Deep learning of topological phase transitions from
		entanglement aspects for two-dimensional chiral p-wave superconductors}
	\author{Ming-Chiang Chung$^{1,2,3}$}\email{mingchiangha@phys.nchu.edu.tw} 
	\author{Tsung-Pao Cheng$^{1}$} 
	\author{Guang-Yu Huang$^{1}$}
	\author{Yuan-Hong Tsai$^{4,5}$}\email{yhong.tsai@gmail.com},
	\affiliation{$^1$ Physics Department, National Chung-Hsing University, Taichung, 40227, Taiwan}
	\affiliation{$^2$ National Center for Theoretical Sciences, Physics
		Divison, Taipei,  10617, Taiwan  }
	\affiliation{$^3$ Physics Department, Northeastern university, 
		360 Huntington Ave., Boston, Massachusetts 02115, U.S.A. } 
	\affiliation{$^4$ AI Foundation, Taipei, 106, Taiwan}
	\affiliation{$^5$ Taiwan AI Academy, Taipei, Taiwan}
	
\begin{abstract}
Applying deep learning to investigate topological phase transitions (TPTs) becomes a useful method due to not only its ability to recognize patterns but also its statistical excellency to examine the mount of information carried	by different types of data inputs. Among possible data types, entanglement-related 	quantities, such as Majorana correlation matrices (MCMs), one-particle entanglement spectra (OPES), and entanglement eigenvectors (OPEEs), have been proved effective, however, are to date mostly restricted to one dimension. Here, we propose practical input data forms based on those quantities to study TPTs and to compare the efficiency of each form on classic two-dimensional chiral $p$-wave superconductors via the deep learning approach. First, we find that different input forms, either matrices or tensors both originated from real MCMs, can affect the precise locations of the predicted transition points.  Next, due to the complex nature of OPEEs, we extract three spatially dependent quantities from OPEEs, one related to the ``intensity'', and the other two related to ``phases'' of particle and hole components. We show that similar to taking OPES directly as inputs, solely using ``intensity'' quantity can only distinguish topological phases from trivial ones, whereas using either whole MCMs or complete OPEE-extracted quantities can provide sufficient information for deep learning to distinguish between phases of matter with different $U(1)$ gauges or Chern numbers. Finally, we discuss certain characteristic features in the deep learning approach and, in particular, they reveal that our trained models indeed learn physically meaningful features, which confirms the potential use even at high dimensions.     
\end{abstract} 

\date{\today}
\maketitle 

\section{Introduction}
Topological matters (TMs) and topological phase transitions (TPTs) have been one of the main research topics for the last 20 years\cite{TMTPT}. Different from conventional phase transitions,  TPTs occur without any broken symmetry. Since the finding of integer quantum Hall \cite{IQH}  as an example of TPT, various TMs discovered theoretically can serve as candidates for TPTs, however, only some of them have been confirmed experimentally. One of the most interesting TMs performing TPTs is two-dimensional chiral $p$-wave superconductors (2D C$p$-SCs)\cite{Volovik,ReadGreen}. In particular, a 2D C$p$-SC has a phase described by the complex superconducting gap function, $\Delta({\bf k}) \sim k_x+i k_y$,  where its phase vector goes around some axis, winding in a clockwise or counterclockwise manner, on the Fermi surface of the material. %For a two-dimensional chiral superconductor, this superconducting gap function can be more explicitly expressed as $\Delta(k) \sim k_x+i k_y$ in the continuous limit. 
Such kind of system possesses trivial and topological phases of matter distinguished by Chern numbers \cite{Schnyder} in the bulk states. Or, alternatively, the bulk-edge correspondence of the 2D topological superconductors also occurs here: for topological phases, gapless chiral edge modes or Majorana edge modes \cite{ReadGreen} can propagate along an geometrical edge. Therefore, observation of such chiral edge states is often, in turn, considered as a signature for the existence of topological C$p$-SCs\cite{SatoReview}. 

Actually the topological properties we mentioned above can also be detected by quantum information related quantities, such as one-particle entanglement spectra (OPES) and entanglement eigenvectors (OPEEs) obtained by diagonalizing the so-called block correlation matrix (BCM) \cite{BCMReview, BCMa, BCMb, BCMc, BCMd}. The simplest example is the Majorana zero modes for one-dimensional $p$-wave superconductors (1D $p$-SCs) that appear as $1/2$ value in the OPES, and their corresponding Majorana edge states also exist in OPEEs \cite{hatsugai06,chung16}. Similarly, signatures such as gapless chiral edge modes in 2D systems would appear in the entanglement spectra, and moreover the entanglement eigenvectors also include those from the chiral edge modes (Majorana edge modes) \cite{hatsugai06}. Therefore, BCM, OPES, and OPEEs can serve as potential candidates of the quantum information related quantities for investigating topological properties of interested systems. Even though we can calculate all these quantities to find out the topological phase transition points, they are very much time consuming, especially for {\it higher} dimensions. And thus one needs to develop new tools to study TPTs, from efficiency perspective. 

In recent years, the neural network (NNW)-based machine learning (ML), namely, deep learning (DL) has drawn a lot of attention in physics community. Due to its data-driven nature, a well-trained NNW model can learn to represent or encode each data point in a given large data set in terms of a more compact vector in internal (hidden) dimensions. DL thus can be efficiently applied for several types of tasks, such as approximating quantum wave functions \cite{Carleo17,Gao17,Deng17a,Deng17b,Nomura17,Kaubruegger18,Glasser18,Choo18,Melko19,Ohtsuki20}, assisting quantum simulations \cite{Arsenault14,Arsenault15,Broecker17,Ryczko19,Sellier19,Suwa19}, and detecting phases of matter \cite{Nieuwenburg17,Carrasquilla17,Ohtsuki16,wang16,Tanaka17,Wetzel17,Hu17,Broecker17b,Chng18,Liu18,Scheurer19, Scheurer20}. In particular, DL has be shown not only to help recognize conventional, symmetry-breaking phase transitions but also to discover non-local, topological ones \cite{kim17a, kim17b,Zhang18,Sun18,Carvalho18,Ming19,Caio19,Greplova20,Zhang21}. In fact, when feeding in appropriate input data forms, NNW models with certain explainable tools may even shed some light on what core ``concepts'' they have learned after training \cite{Zhang20,tsai20,Dawid20}.    

\begin{figure}[t]
	\begin{center}
		\includegraphics[width=7.5cm]{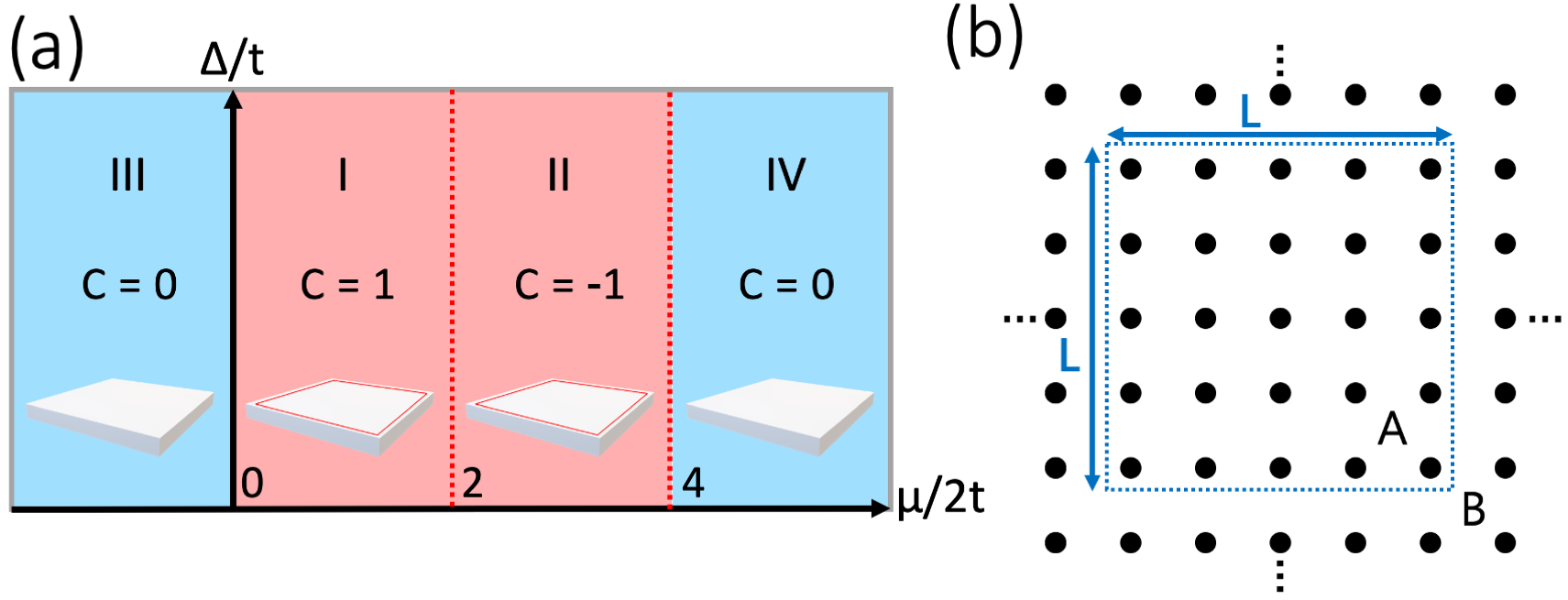}
		\caption{(a) Topological phase diagram
			of 2D Cp-SC, where square-plate inset shows whether
			there exist chiral edge states or not. Phase I and
			II are topological phases with Chern number $1$ and
			$-1$ respectively. Phase III and IV are
			trivial phases with Chern number 0. (b) The infinite 2D square lattice is
			divided into a finite system A with $L\times L$ sites
			and an environment B. }
		\label{fig:model}
	\end{center}
\end{figure}

In this paper, we extend our previous work on the 1D $p$-SCs \cite{tsai20} to 2D C$p$-SCs by using supervised learning to encode quantum information-related quantities such as Majorana correlation matrices (MCMs), BCM-generated OPES and OPEE for recognizing topological phase transitions. Different from 1D systems, there are challenges from relatively complex input data forms to appropriate NNW architecture design for two spatial dimensions. We here propose a feasible way to represent our raw data and a suitable model architecture to analyze the system. Our results are shown to be similar to those in 1D $p$-SCs and the phase transition points are also predicted with good precisions. This significant extension demonstrates the applicability of DL to {\it higher} dimensional topological systems as an efficient new tool. Moreover, as one of the main goals in our series of works, it also encourages researchers to push forward the territory of DL approaches for analyzing more intricate topological materials or even interacting ones. The rest of the paper is organized as follows. In Sec. II, we first define the 2D C$p$-SC model and its related entanglement-based quantities. To apply DL approach, we explain our data preparation in Sec. III, followed by showing our model architecture design for DL in Sec. IV. In Sec. V we then present our results and finally conclude our work with discussions in Sec. VI. 

\section{Model}
The Hamiltonian of the spinless fermion model for a 2D C$p$-SC \cite{TMTPT}  on a square lattice is defined as follows
\begin{equation}
\begin{split}
H = & \sum_{m,n} -t \left(c_{m+1, n}^{\dagger} c_{m,n} +
\mbox{h.c.}  \right)  - t\left(c_{m, n+1}^{\dagger} c_{m,n} +
\mbox{h.c.}  \right)   \\
&  - (\mu - 4t) c_{m,n}^{\dagger} c_{m,n} +  \left(\Delta c_{m+1, n}^{\dagger} c_{m,n}^{\dagger} +
\Delta c_{m+1,n} c_{m,n}   \right)\\  
& + \left( i \Delta
c_{m, n+1}^{\dagger} c_{m,n}^{\dagger}  -
i \Delta  c_{m+1,n} c_{m,n}   \right) ,  
\end{split} \label{H:lattice}
\end{equation}
where $t$ is the nearest-neighbor hopping amplitude, $\mu$ is the on-site chemical potential, and $\Delta$ represents the superconducting pairing potential, which is considered to be real. A complex superconducting pairing potential can be gauge transformed to a real number. With the translational invariance, the lattice Hamiltonian (\ref{H:lattice}) can be Fourier transformed to 
\begin{equation}\label{H:R}
H = -\sum_{\vec{k}} \left (c_{\vec{k}}^{\dagger}, c_{-\vec{k}} \right ) 
\left[ {\mathbf R}(\vec{k}) \cdot{\boldsymbol \sigma} \right] 
\left (c_{\vec{k}}, c_{-\vec{k}}^{\dagger} \right )^T.
\end{equation} 
In Eq.(\ref{H:R}) $\vec{k} = (k_x,k_y)$, ${\boldsymbol \sigma}$ is referred to a vector composed of Pauli matrices, {\it i.e.} $ {\boldsymbol  \sigma} = (\sigma_x, \sigma_y,\sigma_z)$ and ${\vec{R(\vec{k})} =
	(\Delta \sin{k_y}, -\Delta \sin{k_x}, \epsilon_{\vec{k}}) }$ with $\epsilon_{\vec{k}} = 4t - \mu-2t(\cos{k_x} +
\cos{k_y})$.  The complex superconducting potential is momentum dependent, {\it i.e.}, $\Delta(\vec{k}) \sim \sin{k_x} + i  \sin{k_y}$. In the continuous limit it looks exactly the same as  $k_x+ik_y$. 

%Energy spectra, phase transition points 
The quasi-particle spectrum of $H$ can be easily computed as
\be \label{eqn:R}
E_{\pm} =  \pm 2 \left| {\vec{R}} \right| 
= \pm 2\sqrt{\epsilon_{\vec{k}}^2 + \Delta^2(\sin^2{k_x} +
	\sin^2{k_y})}.
\ee
As a result, in the low-temperature limit, the system is basically gapped at any $\mu$ and finite $\Delta$ except at certain phase transition points. Taking $t$ as our energy units, we define two dimensionless parameters $\tilde{\mu} = \mu/2t$ and $\tilde{\Delta} = \Delta/t$ at convenience. In terms of these parameters, one can find that the gap is closed at three places: $\tilde{\mu} = 0, 2,4$ with $(k_x, k_y)=(0,0), (\pi,0)$  or $(0,\pi), (\pi,\pi)$, respectively. By calculating the Chern number of ${\vec{R}(\vec{k})}$,  the ranges $\tilde{\mu} < 0$ and $\tilde{\mu} > 4 $ correspond to trivial superconducting phase with Chern number 0, while the ranges $0 <\tilde{\mu} <2$ and $2 < \tilde{\mu} <4$ correspond to topological superconducting phases with opposite chiralities, {\it i.e.}, with Chern number $1$  and $-1$, separately [see Fig.~\ref{fig:model}(a)]. 

%chiral edge states and localization  length 
According to the bulk-edge correspondence, the topological nature in a bulk system would signify the existence of edge states, when it has an open boundary or a domain wall of the chemical potential $\mu$. For instance, suppose that a domain wall of $\mu(x)$ is put in the $x$-direction of the chiral p-wave superconductor, where $\mu(x) = -\mu_0$ for $x<0$, and $\mu(x)=\mu_0$  for $x>0$. Since the translational invariance is not broken in $y$-direction, $k_y$ is still a good quantum number. In the low momenta limit, one can obtain the edge state as 
\be \label{Eq:decay_func}
\mid \Psi_{k_y}(x,y) \rangle = e^{i k_y y} \exp{\left( -\frac{1}{2\Delta}
	\int_{0}^x \mu(x') dx' \mid \phi_0 \rangle \right)},
\ee 
where $|\phi_0\rangle$ denotes a constant spinor state with energy 
\be
E = -2\Delta k_y. 
\ee
The linear energy spectrum of edge states for small momenta is just one important signature of the chiral edge states. Note that the superconducting gap potential is in the dominator of the decay function of the edge state in Eq.(\ref{Eq:decay_func}), and one can further calculate the  localization length of the edge state as 
\be \label{Eq:coherentL}
\xi = \frac{\vec{v}_F}{2\Delta},
\ee
where $\vec{v}_F$ is the Fermi velocity. Clearly, for large $\Delta$ the Majorana edge states decay much faster than those with small superconducting gap. In fact, it may cause some observation problem for the machine, as we will discuss later in Sec. VI. 

%Entanglement measurement and entanglement spectra
As mentioned in the introduction, using quantum entanglement related quantities is another way to study topological systems. Given a bipartite quantum state $|\Psi_{A  \cup B} \rangle $, a common entanglement measurement is the von Neumann entropy or the entanglement entropy (EE) of a subsystem $A$ : $S_A = - {\tr} \rho_A \log_2 \rho_A$, where $\rho_A$ is the reduced density matrix defined as $\rho_A = \tr_B | \Psi_{A \cup B} \rangle \langle \Psi_{A \cup B}|$. Such quantity indicates that under local operations and classical communications (LOCC), an entangled state with that EE can only be transformed in to a state with the same or lower entanglement quantities \cite{Wooters98}. Therefore, EE can be used as a measurement of entanglement. Note that, however, EE is simply a compact form of quantum entanglement, and can be derived from the eigenvalues of a reduced density matrix.  More explicitly, for a quadratic Hamiltonian, $\rho_A = \bigotimes_k \left[\begin{matrix} \lambda_k & 0\\ 0 & 1-\lambda_k \end{matrix}
\right]$, where $\lambda_k$ are the eigenvalues of the correlation matrix (CM) $C_{i,j} = \tr \rho {\hat{\bs c}}_i \hat{{\bs c}}_j^{\dagger} $ with $\hat{{\bs c}}_i \equiv (c_i, c_i^{\dagger})^T $ and $i,j$ being sites of the subsystem $A$. $\lambda_k$s are known as one-particle entanglement spectra (OPES). In the Fourier space, CM  is a $2\times 2$ matrix for the Hamiltonian (\ref{H:R}) 
\be \label{CM}  
C({\vec{k}}) = \frac{1}{2} \left[1-\frac{{\bs R}({\vec{k}}) \cdot {\bs \sigma} }{R({\vec{k}})} \right], 
\ee 
where ${\bf k}  \in {\bf T}^d$,  {\it i.e.}  each {\bf k} lies on a $d$-dimensional torus,  for a $d$-dimensional  system. Eq.(\ref{CM}) reveals two results: First of all, $C_{i,j}$ commutes with $H_{i,j}$  of  the full Hamiltonian (\ref{H:lattice}). That means, if a block CM (BCM) $C_{i,j}$, where  $ i,j $ are within the subsystem $A$, is diagonalized, the eigenvalues are related to the eigenvalues of  $H_{i,j}$ of the subsystem up to a shift $1/2$ and a normalization factor. It is straightforward to see that the eigenvalues of
$C_{i,j}$ are constrained between $0$ and $1$. Secondly, the eigenvectors of BCM are basically the same as the original Hamiltonian with open boundaries. 

Consequently, on the one hand, the zero energy of the block Hamiltonian for a 1D $p$-SC would correspond to the eigenvalue $1/2$ in a BCM. These Majorana edge states in an open 1D $p$-SC are thus mapped to the maximally entangled states, {\it i.e.}, the edge states in a BCM. On the other hand, the entanglement spectra of a 2D BCM are related to the chiral energy spectra of the block 2D $p$-SC in a different way, compared to the 1D case. One still has chiral entanglement spectra, however, they are not just the same as the chiral energy spectra but exponentially twisted between $0$ and $1$ [see Fig.~\ref{fig:ES}(a)]. Unlike 1D $p$-SCs, there are no OPES with the value $1/2$ except at the point $\mu = 4t$. The chiral edge states of the block 2D $p$-SC also appear, correspondingly, in the entanglement eigenvectors. For a 2D system with square lattice as the interested subsystem,  the chiral Majorana edge states look like a city-wall, as can be seen in Fig.~\ref{fig:OPEE}(a). 

\section{Data Preparation}  
\label{sec:DP}
Quantum information of the system may serve as a good tool to diagnose topological phase transitions,
and especially, the whole information about the focused quantum state is encoded in the entanglement correlations. Based on this fact, in this paper, two common correlators are considered for machines to learn. Considering a 2D lattice of infinite size divided into a finite subsystem $A$ and an
environment $B$ [see Fig.~\ref{fig:model} (b)], we first consider the Majorana correlation matrix (MCM)  in terms of two Majorana fermions defined as $d_{2m-1,2n-1} =- i (c_{m,n} - c_{m,n}^{\dagger})$  and   $d_{2m,2n} = c_{m,n}+c_{m,n}^{\dagger}$ at different sites within the subsystem $A$: $\mbox{MCM}_{{(m,n)},(m',n')}\equiv i \tr \rho_0 d_{2m-1,2n-1} d_{2m',2n'}  = \tr \rho_0 (c_{m,n} -c_{m,n}^{\dagger})( c_{m',n'} + c_{m',n'}^{\dagger}) $, where $\rho_0$ represents the density matrix of the ground state. Second, as mentioned in the last section, the block correlation  matrix (BCM) for subsystem $A$ is defined as $\mbox{BCM}_{(m,n),(m',n')} = \tr\rho_0 \hat{\vec{c}}_{m,n} \hat{\vec{c}}_{m',n'}^{\dagger}$ with $\hat{\vec{c}}_{m,n} \equiv (c_{m,n}, c_{m,n}^{\dagger})^T$ and $(m,n), (m',n') $ representing 2D sites of the block $A$. The exact matrix elements of MCM and BCM are listed in Appendix. By diagonalising BCM, one obtains the entanglement spectra $\lambda_k$ and the entanglement eigenvectors. In order to recognize topological phase transitions in the 2D C$p$-SC, we next explicitly explain the three types of data forms used for the deep learning processes. 

Let us consider a subsystem $A$ of size $L \times L$ embedding in an infinite square lattice: (i) The first data type fed into machine (NNW) is the MCM, which can have two different forms: One form is
a $L^2 \times L^2$ matrix, where a 1D-like system of size $L^2$ is formed by flattening the 2D square lattice into 1D. The other form is a $L \times L \times L \times L$ tensor. Since the entries in MCMs are real (see Appendix), they can be easily handled by DL. 
(ii) The second data type is OPES which can be obtained by diagonalizing BCMs. They are all $2 L^2$-dimensional vectors. (iii) The third data type is composed of OPEEs, which are also obtained by diagonalizing BCMs. Arranging OPEEs to a proper form for DL training is very cumbersome due to the fact that the components in each OPEE are complex numbers. Additionally, each eigenvector is a $2 L^2$-dimensional vector with the factor of 2 standing for Nambu space. In other words, for the site $i$, each eigenvector has both particle and hole components represented as complex $Z_i^{p}$ and $Z_i^{h}$. Therefore, due to particle-hole symmetry, one can simply take three necessary quantities for representing information of each OPEE at site $i$: $r_i^2 =( r_i^p)^2 + (r_i^h)^2$, $\tan \theta_i^p$, and $\tan \theta_i^h$, where $r_i^p$ ($r_i^h$) and $\theta_i^p$ ($\theta_i^h$) are the absolute value and the angle of $Z_i^p$ ($Z_i^h$), respectively. From them one can then form tensors with dimension  $6 L^2 \times L \times L$ and feed them into machine (NNW) for DL training. 

\begin{figure}[t]
	\begin{center}
		\includegraphics[width=8.5cm]{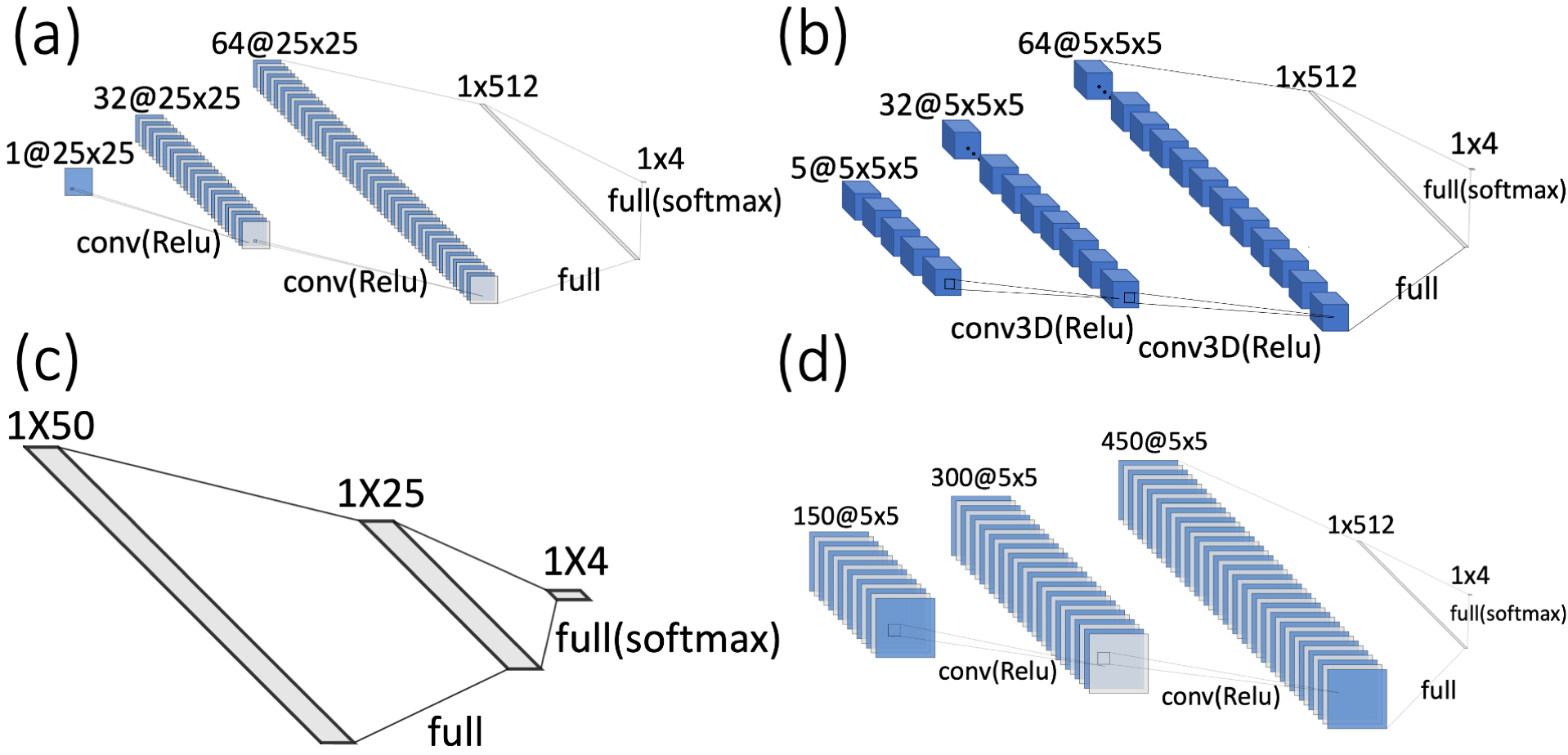}
		\caption{The schematic illustrations of the convolutional
			neural network used in this work for (a) MCM of matrix forms  (b) MCM
			of tensor forms (c) OPES (d) OPEE.}  \label{fig:DLArchitecture}
	\end{center}
\end{figure}

\section{Deep-Learning Approach} \label{sec:DLA}
In the previous study on 1D $p$-SCs \cite{tsai20} we used different model architectures to find the phase
transition boundaries, including deep neural networks (DNNs) and convolutional neural networks (CNNs), according to data forms we considered. In particular, CNNs are considered to be efficient at recognizing patterns of image-like data due to their inherent inductive biases such as translational equivariance and locality. As matrices and tensors can be interpreted as one- or multi-channel images, employing CNN-like model architectures would be our first choice in the DL approach. However, different from 1D cases, recognizing phase transitions in 2D C$p$-SCs are rather challenging for DL due to complicated structures of the data, which are naturally multi-channel (when viewed as images) and complex. Therefore, the design of model architectures for DL may have to vary depending on the data forms we adopt. 

We handle the data of MCMs in two ways as discussed in Sec. III. The way to treat MCMs in a matrix form is similar to that used for 1D systems where a CNN with 2 convolutional layers are employed. Explicitly, we build a deep CNN using the PyTorch framework \cite{pytorch}, as shown in Fig.~\ref{fig:DLArchitecture}(a): Begin with the convolutional module, and then connect to fully connected layers. The convolutional module is composed of two consecutive convolutional layers with both filters of kernel size $l \times l$ and ReLU activation functions \cite{GBC}, where $l$ is adjustable and chosen from 3 to 7 according to the system size $L$. The number of filters is 32 for the first layer, and 64 for the second one. The zero-padding technique \cite{GBC} on the input data is used before each convolution to keep the spatial size intact and no pooling layers are inserted due to
our small image spatial size. After convolutions, the output is put into a classifier consisted of a fully-connected ReLU activated layer with 512 neurons and a fully-connected four neuron softmax layer \cite{GBC}. The final output then can be interpreted as the probabilities for each one of four phases shown in Fig.~\ref{fig:model}(a). 

The other way is to treat MCMs as a tensor form. We use a three-dimensional CNN to find the phase boundaries, as demonstrated in Fig.~\ref{fig:DLArchitecture}(b). Due to the fact that a $L\times L \times L \times L $ tensor can be reshaped as $L$ three dimensional tensors of the shape $L \times L \times L$, we can build a similar model architecture like the previous one, while the convolutions become three-dimensional, with learnable filters of kernel size $3 \times 3 \times 3$ instead of $3\times 3$. 

On the other hand, as long as OPES and OPEE of BCMs are concerned, we simply consider these model architectures separately: For OPES it is easier since they are vectors, and we can employ DNN architecture for DL training, as shown in Fig.~\ref{fig:DLArchitecture}(c), while for OPEE, it becomes much more complicated. Diagonalizing a BCM gives rise eigenvectors (with complex numbers in general) which can be arranged as a matrix with shape  $2L^2\times 2L^2$. As mentioned in Sec. III, we extract 3 real numbers from each eigenvector: $r_i^2, \theta_i^p$, and $\theta_i^h$ for each site $i$. Namely, from each eigenvector, three vectors with $L^2$ dimensions are generated. After reshaping each $L^2$-dimensional vector into a $L\times L $ matrix according to the spatial relations in the subsystem $A$, we actually obtain $6 L^2$ matrices of size $L\times L$. Taking $L=5$ as an example, we build a two-dimensional CNN architecture as shown in Fig.~\ref{fig:DLArchitecture}(d) for this input data form. In other words, one can view the data form as a 150-channel $5\times 5$ ``image'' and it is fed into the model with two convolutional layers at the beginning. The number of filters chosen is 300 for the first layer and 450 for the second one.  The other part of the model architecture would be the same as usual [see Fig.~\ref{fig:DLArchitecture}(d)]. 

Since we adopt supervised learning strategy to train our model, the ``labeled'' training data are chosen at four $\tilde{\mu}$ points, each of which is expanded within a window of size $0.1$ in the unit of $\tilde{\mu}$. They simply correspond to four possible phases deep inside the phase diagram [see Fig.~\ref{fig:model}(a)]. Setting the train-validation split ratio as 0.2, the optimization for training is then performed by ADAM algorithm at learning rate $10^{-3}$ with cross entropy as the loss function \cite{Kingma15}. Once the loss after training is converged, at the inference stage we fix whole parameters in the trained model and feed with new data for prediction.

\begin{figure}[t]
\begin{center}
\includegraphics[width=8.5cm]{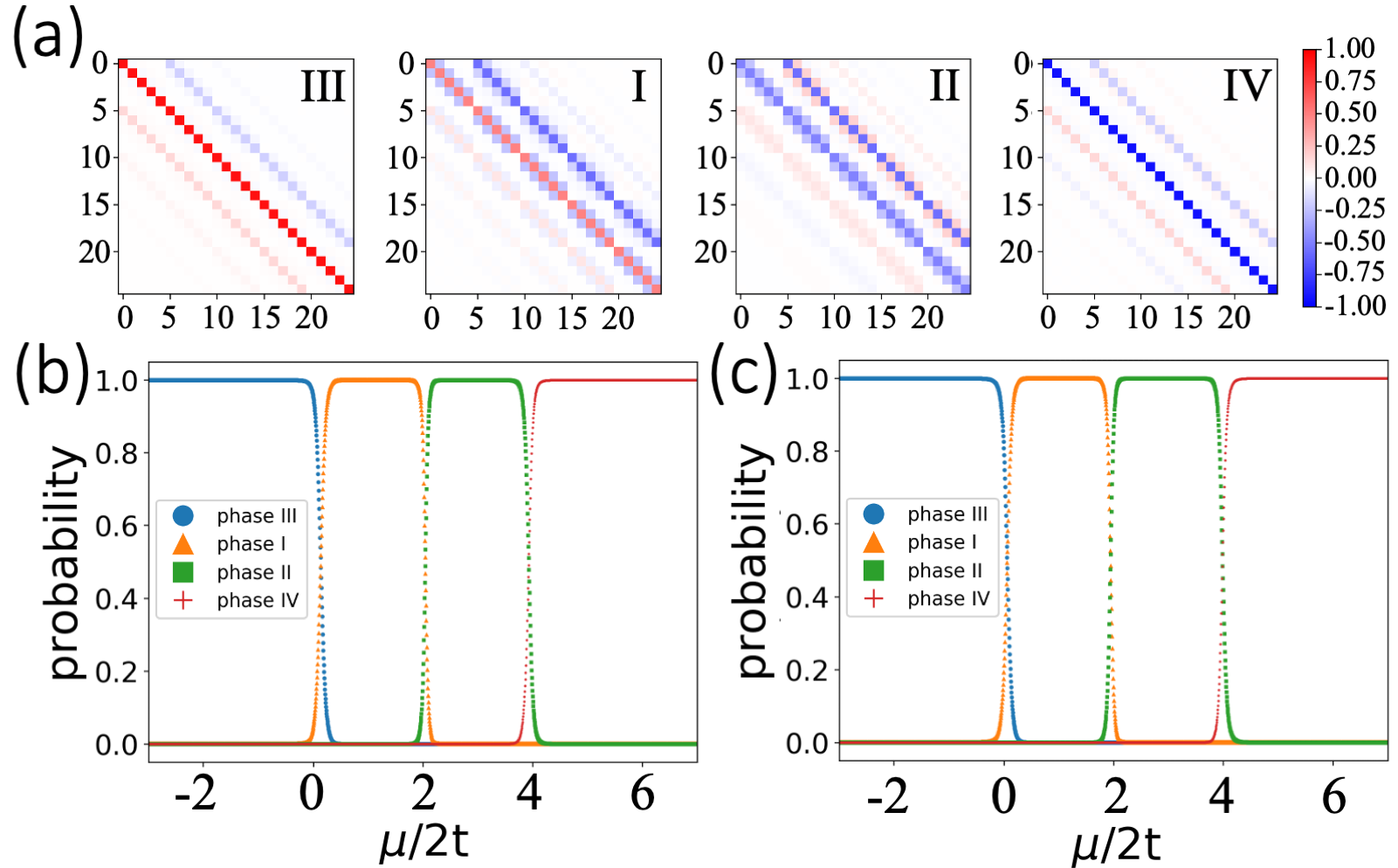}
\caption{(a) Matrix elements of MCM shown as images  for $\mu/2t = -1, 1, 3, 5$ representing phase III, I, II, IV respectively. Here the system size is chosen as $5\times 5$.  (b) Each neuron output of the final softmax layer with matrices as inputs, corresponding to the probability of each phase, as a function of $\mu/2t$ (unseen data) with $\Delta/t = 1,  L=5$. (c) same	data set as (b) with the only difference: Here tensors are fed into DL  instead of matrices. For (b) and (c) the training sets from MCMs are prepared at around $\mu/2t = -10, 1, 3, 14$ of a window width $0.1$.  }
\label{fig:MCMResult}
\end{center}
\end{figure}

\section{Results}
\subsection{DL from MCM-generated data}
%According to the aforementioned data preparations we can feed different types of entanglement-related quantities of 2D chiral $p$-wave SC into DL and  the main goal is to study two dimensional chiral topological phase transitions. 
By calculating a thousand of MCMs for a subsystem of size $L\times L$, embedded in a full system under periodic boundary conditions, around a given $\tilde{\mu}$ for each phase in the phase diagram, we prepare to generate the training labeled ``images''. Note that, for simplicity, we choose $\tilde{\Delta} =1$ for the training set. The first input data form is the elongated $L^2\times L^2$ matrix as discussed in Sec. III.  As shown in Fig.~\ref{fig:MCMResult}(a) the ``images'' of correlations vary for different $\tilde{\mu}$ values. 

Similar to 1D $p$-SCs, our trained model can distinguish different phases for a given dataset and discover the phase boundaries by feeding a set of unseen data points along $\tilde{\mu}$. Explicitly, this can be seen in Fig.~\ref{fig:MCMResult}(b): The probability of the neuron output for predicting phase III drops from 1 at $\tilde{\mu}=-2$ to 0 at $\tilde{\mu}=0.5$, whereas another output for predicting the phase I arises from 0 to 1. These two curves cross each other at $\tilde{\mu}^{\star} = 0.132, 0.087, -0.062, 0.057$ for subsystem size $L=5, 6, 7, 8$, respectively, as listed in TABLE~\ref{tab:CPs}. These probabilities indicate that the trained CNN model indeed recognizes the occurring of a phase transition. Similarly, the model also realizes two other phase transition points at $\tilde{\mu}^{\star} = 2.042, 2.067, 1.987, 1.982$ and $\tilde{\mu}^{\star} = 3.922, 3.947, 3.947, 3.952$ for $L=5,6,7,8$, respectively, to distinguish phases I and II, and phases II and IV.  

The second type of input data form refers to $L\times L \times L \times L$ tensors, which can be viewed as $L$-channel, 3D ``images''. When fed into our trained 3D CNN model, the probability curves as a function of $\tilde{\mu}$ are similar to those of matrix data form just mentioned above, as shown in Fig.~\ref{fig:MCMResult}(c). However, the predicted critical points are located at slightly different places, for instance, $\tilde{\mu}^{\star} =0.057, 1.942, 3.982$ for $L=5$.  More results for various $L$ can be seen in TABLE~\ref{tab:CPs}.

Comparing the results of matrix form from MCMs with those of tensor form, it is clear that for smaller sizes, for instance $L=5$, the tensor form provides more accurate results than the matrix one. This observation might be understood as follows. When preparing the input data to an appropriate form, the matrix form often has issues of artificial boundaries caused by rearranging original square lattice to a line, while the tensor one preserves the geometry of the two-dimensional system. However, the advantage of the tensor form vanishes for larger sizes, {\it e.g.} $L=8$, as the portion of the artificial boundaries in the whole input data diminishes and thus becomes less relevant. 

\begin{table}[h]
	\centering
	\begin{tabular}{|l|c|c|c|c|c|l|}\hline
		& \multicolumn{3}{c|}{Matrix Inputs} & \multicolumn{3}{c|}{Tensor Inputs}\\
		& 1st CP  & 2nd CP & 3rd CP    & 1st CP & 2nd CP & 3rd CP  \\\hline
		$L=5$  & 0.132 & 2.042  & 3.922 & 0.057  &1.942  &3.982    \\
		$L=6$ & 0.087 & 2.067 & 3.947 & -0.012 & 1.962 & 3.947\\
		$L=7$ & -0.062 & 1.987 & 3.947 & 0.017 & 1.957 & 3.957 \\
		$L=8$   & 0.057 & 1.982 & 3.952 & 0.057 & 1.997  & 3.937  \\\hline
	\end{tabular}
	\caption{Predicted critical points by training  MCMs for different
		sizes $L=5,6,7,8$. }
	\label{tab:CPs}
\end{table}

\subsection{DL from BCM-generated data}
We next consider quantum-information related quantities generated from BCMs as our input data. As discussed in Sec. III, for a focused subsystem of size $L\times L$ (totally $L^2$ sites) embedded in a large system, there are two related quantities which can serve as input ``images'': (i) The eigenvalues of BCMs, namely, the so-called one-particle entanglement spectra (OPES), and (ii) the eigenvectors (OPEEs) of BCMs. 

\begin{figure}[t] 
	\begin{center}
		\includegraphics[width=7cm]{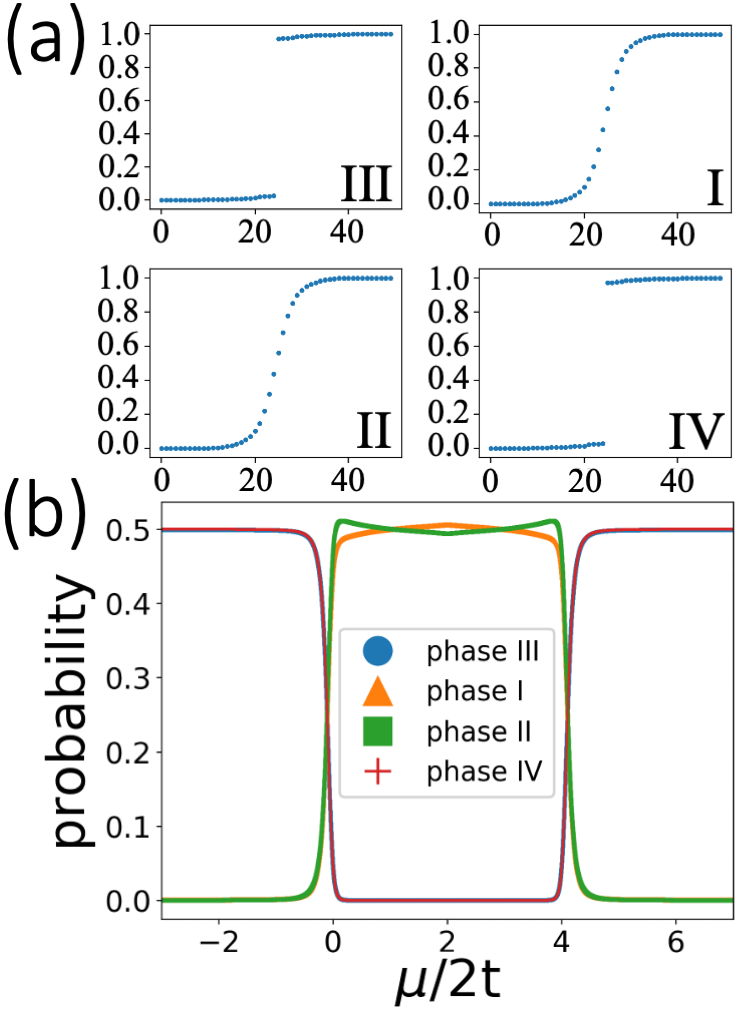}
		\caption{(a) OPES for phase III, I, II and IV. For topological phases I and II, there exist gapless chiral spectra, while for trivial phases III and IV the spectra are gapped.  (b) The neuron output phase diagram of OPES as inputs is shown as a function of $\mu/2t$. The training sets from the OPES are prepared at around $\mu/2t = -10, 1, 3, 14$ of a window width $0.1$. The phase boundaries between topological phases  and trivial ones are clear, however, the probabilities corresponding the whole four phases are near $1/2$, suggesting either topological phases I and II or trivial phases III and IV confuse the neural network.} 
		\label{fig:ES}
	\end{center}
\end{figure}	

The entanglement spectra of the 2D-C$p$SCs are shown in Fig.~\ref{fig:ES}(a). It is obvious that for topological phases, {\it i.e.}, $0 \le \tilde{\mu} \le 4$, there exist chiral spectra near $\lambda_k = 0.5$, whereas for trivial phases most entanglement spectra are near $0$ or $1$ without the values near $1/2$.  By taking the same training procedure and waiting until the training and validation losses converge, the final inference for unseen data at $\tilde{\Delta} =1$ as a function of $\tilde{\mu}$ are shown in Fig.~\ref{fig:ES}(b). For the topological phases $0 \le \tilde{\mu} \le 4$, {\it i.e.}, phases I and II  cannot be distinguished and the predicted probability are both near $0.5$, indicating that the network is confused by these two phases due to their similar eigen-spectra. The same situation also happens between phases III and IV. The lack of ``phase'' information, or equivalently, complex nature revealed  in OPES may account for such disability.
%Physically, phase I and phase II belong to the same topological superconducting phase but with different $U(1)$ gauge, so do phase III and IV to same trivial phase. 
In fact, unlike MCMs, OPES obtained from BCMs can only recognize the phase boundaries between topological and trivial phases. The two phase transition points predicted by OPES are $(-0.105, 4.106)$ for $L=5$, $(0.067, 3.933)$ for $L=6$, $(0.097, 3.902)$ for $L=7$ and $(0.063, 3.937)$ for $L=8$, respectively. Regardless of the even-odd site effect, which is common in the solid state physics, the precision of the predicted values for the phase boundaries becomes better through the enlargement of the system. This actually reflects the finite-size effect behind our numerical experiments. 
	
\begin{figure}[t] 
	\begin{center}
		\includegraphics[width=8.5cm]{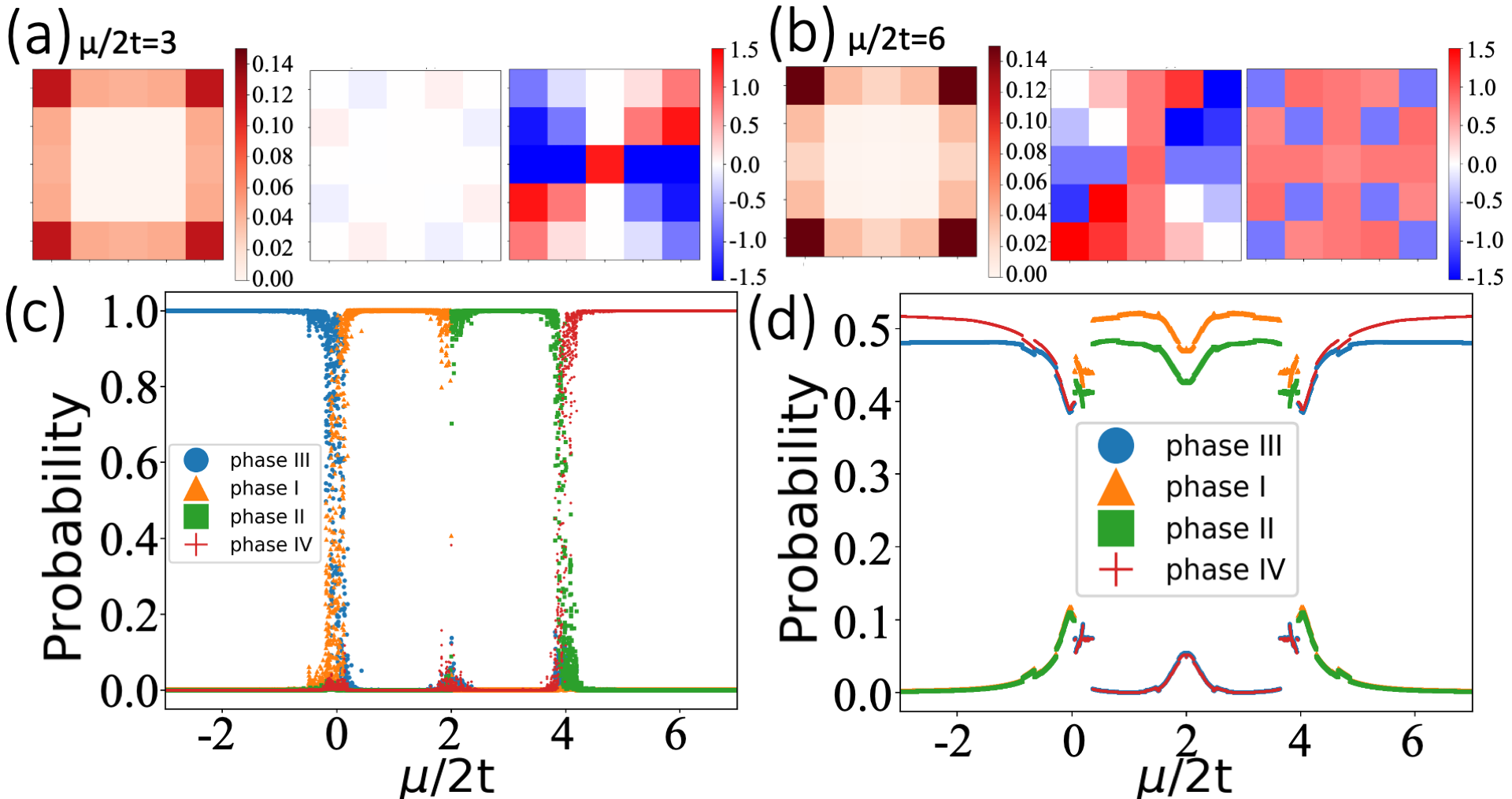}
		\caption{(a) $r_i^2$, $\theta_i^p$ and
			$\theta_i^h $ images of OPEE  for a topological
			phase. Here we choose $\mu/2t=3$. (b) $r_i^2$, $\theta_i^p$ and
			$\theta_I^h $ images of OPEE  for a trivial
			phase. Here we choose $\mu/2t=6$. (c) The neuron
			output “phase diagram” is shown as a function of
			$\mu/2t$ with $\Delta/t=1, L=5$. The training sets from
			whole eigenvectors of BCMs are prepared at the same $\mu/2t$  regions
			mentioned  in Fig.~\ref{fig:ES} and Fig.~\ref{fig:MCMResult}. (b) Similar diagram as (a) but with
			only inputs from images of $r_i^2$. The results are similar to those
			trained by OPES.}\label{fig:OPEE}
	\end{center} 
\end{figure}

The other quantum-information related quantities are OPEEs, which have more abundant structures than those of OPES. We decouple each OPEE into $r_i^2$, $\theta_i^p$ and $\theta_i^h$ for the site $i$. As mentioned in Sec. IV, we further reshape these data to be $6L^2$-channel $L\times L$ ``images'', where the image shape, $L\times L$, corresponds to original spatial information. Fig.~\ref{fig:OPEE}(a) and (b) show the images of $r_i^2$, $\theta_i^p$ and $\theta_i^h$ for $\tilde{\mu} = 3$ (topological phase) and $\tilde{\mu} = 6$ (trivial phase), respectively, at $L^2$th channel, with $\tilde{\Delta} =1$, and the size $L=5$. An important signature to distinguish two-dimensional chiral topological from trivial phases is the existence of chiral Majorana edge modes. This may appear as a wall structure in the $r_i^2$ images, which is clearly seen in Fig.~\ref{fig:OPEE}(a), but is not visible in Fig.~\ref{fig:OPEE}(b). 

%In the training CNN model for OPEEs, we keep the first convolution and then connect with two consecutive residual blocks. And then before the final dense layer-based classifier the global average 2D is used. 
After training a CNN model by feeding in OPEE data, the predicted probability for each phase as a function of $\tilde{\mu}$ at $\tilde{\Delta} = 1, L=5$  is shown in Fig.~\ref{fig:OPEE}(c). Unlike the OPES case, the ability to distinguish phases with different topological sectors (or trivial gauges) has been restored by using OPEEs as input data. The obtained phase transition points are $\tilde{\mu} = 0.000, 2.000, 3.938$, respectively. 

In order to find out necessary components in an OPEE to provide enough phase or gauge information for each phase of matter, we take $r_i^2$ as the sole input data, and show the result in Fig.~\ref{fig:OPEE}(c). It is obvious that, similar to OPES, providing $r_i^2$ data is only capable of finding the phase boundaries between topological and trivial phases, since the probabilities for finding phases I and II, or phases III and IV are all near $50 \%$. Until we add more informations, such as $\theta_i^p$ and $\theta_i^h$, together with $r_i^2$, the whole ability to distinguish various phases can then be recovered by DL. In other words, $r_i^2$ component plays a role to distinguish topological phases from trivial ones, while $\theta_i^p$ and $\theta_i^h$ serve as a key to detailedly distinguish different topological (trivial) phases among themselves that carry different angular or gauge informations.    

\section{Discussion and Conclusion}

\begin{figure}[t] 
	\begin{center}
		\includegraphics[width=8.5cm]{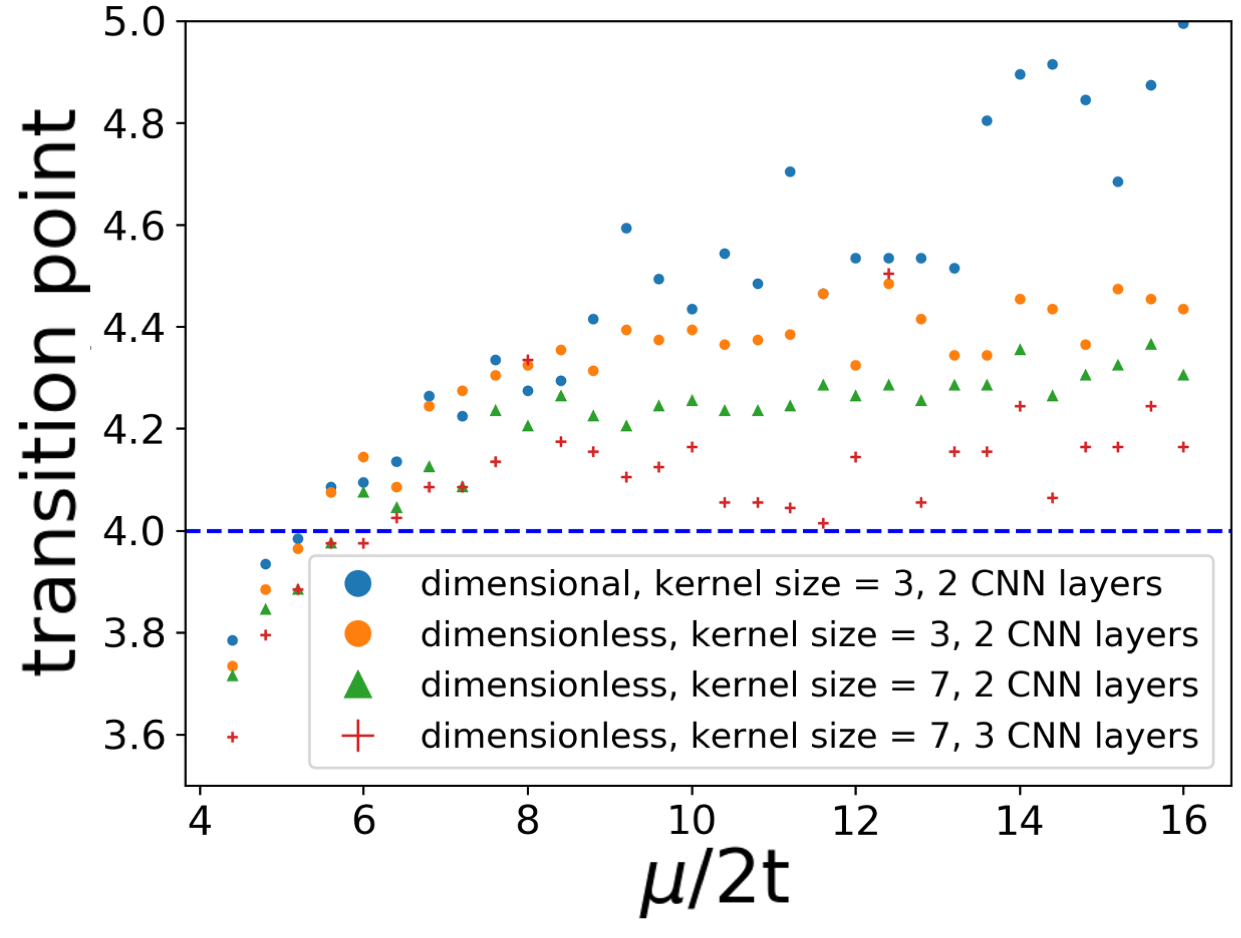}
		\caption{Phase transition points
			predicted by moving the training points $\mu/2t$
			symmetrically  from 0 to -12  for phase III and from 4 to 16 for
			phase IV with $\Delta/t=2$ and
			$L=5$.
			The other training points
			remain the same as mentioned in
			FIg.~\ref{fig:ES}.  The blue solid circles represent
			the results from dimensional coordinates with 3 kernel
			size and 2 CNN layers. The orange circles are those
			from dimensionless analysis with 3 kernel size and 2
			CNN layers. The green triangles shows the results from
			dimensionless coordinates with 7 kernnel size and 2 CNN
			layers. The red crosses represent the outcome from
			dimensionless analysis with 7 kernel size and 3 CNN
			layers. We see that the results are improved by imposing
			different methods. }\label{fig:Delta}
	\end{center} 
\end{figure}

Two issues are worth mentioning here, and they both relate to certain characteristic features in the DL approach. Firstly, when enlarging the superconducting gap $\Delta$,  the  localization length of the edge state [see Eq.(\ref{Eq:coherentL})] decreases inversely with $\Delta$. As a result, if we use the original MCMs as input data, the edge states become more and more difficult to detect by DL because they disappear too fast (along to the direction orthogonal to propagation) to observe. We call it the undressed analysis by dimensional procedure. Therefore, we in turn consider the dimensionless procedure, where the dimensionless coordinates, {\it i.e.}, $\vec{r}_{\Delta} = \vec{r} \Delta$ are utilized. In other words, instead of  $\mbox{MCM}_{{(m,n)},(m',n')}$ we use $\mbox{MCM}_{{( m \Delta ,  n \Delta)},(  m' \Delta,  n' \Delta )}$ as the input data.  Due to the fact that  localization length of a dimensionless MCM is independent of $\Delta$, the predicted phase transition points could be improved. 

In fact, there are other ways to improve the precision of the predicted transition points. For instance, by increasing the kernel size of convolutional layers or by adding more layers. They are both designed to detect more detailed structures of MCMs, and thus they have similar effect on the results. One may use them separately or altogether when designing her/his own model. 

In Fig.~\ref{fig:Delta}, we show the results of such improvements. In order to obtain stable results, we change the training centers for phases III and IV as $4-\tilde{\mu}$ and $\tilde{\mu}$, respectively, and fix those for phases I and II at $\tilde{\mu}$ equal to $1$ and $3$. Note that $\tilde{\mu}$ would serve as the variable of $x$-axis in the figure. The predicted phase transition points increase linearly at the beginning and then saturate after certain $\tilde{\mu}$. The solid blue circles represent the results following dimensional procedures with a kernel size equal to 3 and two convolutional layers, {\it i.e.}, the original treatment. The trend of them often goes higher than the theoretical critical point $\tilde{\mu}=4$ and not stably saturated. The average value of the saturate plateau by using $\tilde{\mu}=10$ as a cutoff is $4.67$. In contrast, the orange solid circles represent the results following dimensionless procedures. Their trend is much more stable than that from taking dimensional procedures, and the average predicted value is $4.41$, which is well improved. Moreover, the green triangles are the results following dimensionless procedure as well, but with 2 convolutional layers and with the kernel size equal to 7. It continues improving the predicted transition point to $4.28$. Finally, we deepen aforementioned model architecture to 3 convolutional layers and hence obtain the red crosses which averagely predict the transition point equal to $4.15$.  Similar effect also happens for ${\tilde{\Delta} <1}$, where the predicted transition points contract by using dimensional coordinates. This can also be fixed by using dimensionless coordinates. 

These kinds of improvements show a characteristic feature in DL approach. One should regard DL as a quasi-experimental setup with a definite precision (once the model is designed). If some pattern can not be observed by DL, it may not mean that DL is unable to find the solutions; instead, it could be the  resolution set by the utilized parameter(s) or model architecture is simply too poor to find the target. One
should improve the physical parameters or change the architecture to re-tune the resolution in order to accomplish the task.  

\begin{figure}[t] 
	\begin{center}
		\includegraphics[width=8.5cm]{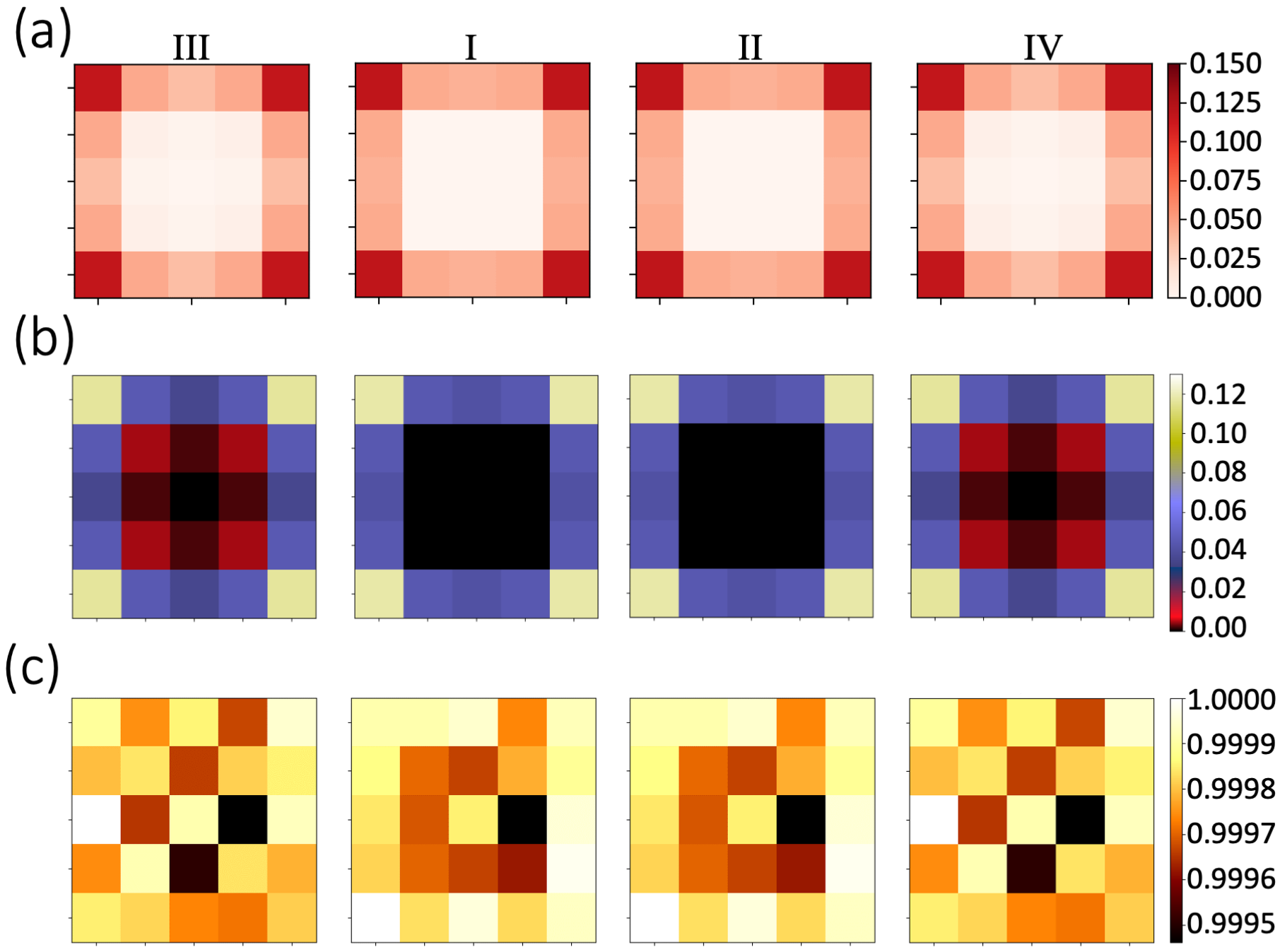}
		\caption{(a) Selected  $L^2$th OPEE
			($\mu/2t = -1, 1, 3, 5$, {\it i.e.} phase III, I, II
			,IV respectively) with  gradually changed color bar, (b) with more distinct color bar and  (c) their corresponding gradient-based class
			activation maps (Grad CAM). One can see  that
			the Grad CAM really finds the features of Majorana
			edge modes for topological phases to distinguish
			them from trivial ones.}\label{fig:gradcam}
	\end{center} 
\end{figure}

The second issue is about the explainability of DL approach, namely, what exactly a DL model sees through the whole black-box-like procedures. One of the most important features about the 2D C$p$-SC is the existence of chiral edge modes in the OPEEs, as we have illustrated in Fig.~\ref{fig:OPEE}(a). This
provides a good chance to examine the feature found by DL. As for the 1D $p$-SC in Ref.~\onlinecite{tsai20}, one can use a gradient-based class activation map, called Grad-CAM \cite{GradCAM}, to visualize the key regions learned by the neural network for a given class. Such a method provides a potential explainable tool for DL. The details of applying this method to the prediction of OPEE ``images'' for $p$-SCs can be found in Ref.~\onlinecite{tsai20}. 

In Fig.~\ref{fig:gradcam}(a), the $r_i^2 $ of the $L^2$th channel ``images'' via OPEEs for phases III, I, II and IV, {\it i.e.}, $\tilde{\mu}=-1, 1, 3, 5$, respectively, are shown  with gradually changed colors.  With bare eyes, they can not be distinguished easily. In order to find the differences between topological and trivial phases, we change the color bar as shown in Fig.~\ref{fig:gradcam}(b).  For topological phases I and II , the chiral edge states can be seen clearly, whereas for trivial ones, the figures are like mosaic, which show no sign of edge states. To justify the explainability of DL via Grad-CAM, Fig.~\ref{fig:gradcam} (b) represents the corresponding CAM results. Remarkably, our trained CNN model with Grad-CAM notices the edges around the boundaries with higher attention weights for predicting the topological phases, while for trivial ones, only mosaic-like patterns of weights appear. In other words, our model actually learns what to focus on and uses chiral edge states noticed to characterize the difference between topological and trivial chiral
$p$-wave superconductors.  

 Using entanglement-related quantities to treat two-dimensional topological systems has several advantages or differences compared to other methods used in several references\cite{kim17a, Sun18, Carvalho18}: (i) Overall we use smaller sizes to obtain higher precision of the phase transition points  due to the reason that MCMs and BCMs contain not only the informations of the subsystem but also of the environment. (ii) We do not use any topological quantities like the Hall conductivity \cite{kim17a} to obtain phase transition points, which makes our method more general to detect not only topological  but also other types of phase transitions. (iii) We work on the real space other than the momentum space to obtain the transition points compared with Ref. \onlinecite{Sun18}. (iv) In none of the aforementioned references  edge modes are observed by using deep learning as in our work. 

In conclusion, we have studied topological phase transitions of 2D C$p$-SCs via DL approach. We primarily consider quantum information related quantities, MCMs, and OPES, OPEEs from BCMs, as source data for machine learning. Due to more complicated nature of those quantities at two dimensions, we propose practical ways to deal with each of them for DL purpose. As to the MCMs, inputs of tensor form give better precision in determining phase boundaries than those of matrix form, particularly for smaller system size due to the artificial boundaries present when reshaping MCMs into a matrix form. On the other hand, although the OPEEs can provide abundant information, they need some special treatment for DL to avoid overfitting. We overcome this obstacle by taking $r_i^2$, $\theta_i^p$, and $\theta_i^h$ in OPEEs as the suitable input form. Unlike using OPES that can only distinguish between topological and trivial phases, both MCMs and OPEEs can further distinguish between different topological sectors or superconducting states with distinct $U(1)$ gauges. This extra distinguish-ability is found to be in $\theta_i^p$ and $\theta_i^h$, as they provide the missing angular or ``phase'' information. To reveal what our trained CNN models have learned from OPEEs, we employ Grad-CAM method and find that they decide whether the phase is topological or not, depending on whether chiral Majorana edge modes are present or not. Our results confirm again the usefulness of DL not only for recognizing topological phase transitions via entanglement quantities and but also for systems at high dimensions. 

\section{Aknowledgement}
M.C.Chung acknowledges the MoST support under the contract NO. 
108-2112-M-005 -010 -MY3 and Asian Office of Aerospace Research and Development
(AOARD) for the support under the award NO. FA2386-20-1-4049. 

\appendix
\section{APPENDIX : CORRELATION MATRICES}
The matrix elements of BCM read
\begin{equation}
BCM_{(m,n),(m',n')} = \begin{bmatrix} \alpha(m,n;m',n') & \beta(m,n;m',n') \\ \gamma(m,n;m',n') &
\delta(m,n;m',n') \end{bmatrix} 
\end{equation}
where 
\begin{equation} 
\begin{split}
\alpha(m,n;m',n') &=  \frac{1}{2} \delta_{m,m'} \delta_{n,n'} - \frac{1}{4\pi^2}
\int_0^{\pi} \int_{-\pi}^{\pi} \frac{dk_x dk_y}{R} \\
&\cos{[k_x(m'-m) + k_y (n'-n)]}  \\
& \times \left[\cos{k_x} + \cos{k_y} +\frac{\mu}{2} -2 \right],
\end{split}
\end{equation}
\begin{equation}
\begin{split}
\delta(m,n;m',n,) & =  \frac{1}{2} \delta_{m,m'} \delta_{n,n'} + \frac{1}{4\pi^2}
\int_0^{\pi} \int_{-\pi}^{\pi} \frac{dk_x dk_y}{R} \\
&\cos{[k_x(m-m') + k_y (n-n')]} \\
&  \times \left[\cos{k_x} + \cos{k_y} -\frac{\mu}{2} -2 \right],
\end{split}
\end{equation}
\begin{equation}
\begin{split}
\beta(m,n;m',n') & = -\frac{\Delta}{4 \pi^2} \int_0^{\pi}
\int_{-\pi}^{\pi} \frac{dk_x dk_y}{R} \\ & \sin{[k_x(m'-m) + k_y (n'-n)]} \\ 
& \times [\sin{k_x} + i \sin{k_y} ],
\end{split}
\end{equation}
and
\begin{equation}
\begin{split}
\delta(m,n;m',n') = & \frac{\Delta}{4 \pi^2} \int_0^{\pi} \int_{-\pi}^{\pi}
\frac{dk_x dk_y}{R} \\ & \sin{[k_x(m'-m) + k_y (n'-n)]} \\ 
& \times [\sin{k_x} - i \sin{k_y} ],
\end{split}
\end{equation}
with Kronecker delta $\delta_{i,j}$ and 
\begin{equation}
R \equiv |\vec{R}|
\end{equation}
as shown in Eq. (\ref{eqn:R}).

MCM can be obtained from BCM in the following way:
\begin{equation}
\begin{split}
MCM_{(m,n),(m',n')} = & \alpha(m,n;m',n') - \delta(m,n;m',n')  \\
& + \beta(m,n;m',n') - \gamma(m,n;m',n').
\end{split}
\end{equation}
Hence the matrix elements read:
\begin{equation}
\begin{split}
& MCM_{(m,n),(m',n')} = \\ &   \frac{\Delta}{4 \pi^2} \int_0^{\pi} \int_{0}^{\pi} \frac{dk_x
	dk_y}{R}   \left[(\cos{k_x} + \cos{k_y} + \mu/2 -2)\right. \\
&  \times\left. \cos{k_x(m'-m)} \cos{k_y (n'-n)}  \right] \\
& + \left[\Delta \sin{k_x} \sin{k_x(m'-m)} \cos{k_y(n'-n)} \right].  
\end{split} 
\end{equation}

Note that the matrix elements of BCMs are complex, and  of  dimension
equal to $2L^2
\times 2 L^2$, whereas those of MCM are real and of dimension equal to $L^2 \times
L^2$. This shows the advantage of using MCMs instead of BCMs as inputs
to feed into DL: they cost much smaller computer space and less time
in the calculations.


\begin{thebibliography}{30}
	\bibitem{TMTPT}  for a review, see B. A. Bernevig with T.L. Hughes{\it Topological Insulators and Topological Superconductors} (Princeton University Press, 2013).
	\bibitem{IQH} K. V. Klitzing; G. Dorda; M. Pepper, Phys. Rev. Lett. {\bf 45}, 494 (1980). 
	\bibitem{Volovik} G. E. Volovik, JETP Lett. {\bf 70}, 609 (1999). 
	\bibitem{ReadGreen} N. Read and D. Green, Phys. Rev. B, {\bf 61}, 10267 (2000). 
	\bibitem{Schnyder}  A. P. Schnyder, S.  Ryu, A Furusaki and A. W. W. Ludwig  Phys. Rev. B {\bf 78 }, 195125 (2008).
	\bibitem{SatoReview} M. Sato and Y. Ando, Rep. Prog. Phys. {\bf 80}, 076501 (2017).
	\bibitem{BCMReview} For a review, see I. Peschel and V.  Eisler,  ``Reduced density matrices and entanglement entropy in free lattice models'', J. Phys. A: Math. Theor. {\bf 42}, 504003 (2009).
	\bibitem{BCMa} M.-C. Chung and I. Peschel, Phys. Rev. B {\bf 64}, 064412 (2001).
	\bibitem{BCMb} I. Peschel, J. Phys. A {\bf 36}, L205 (2003). 
	\bibitem{BCMc} S. A. Cheong and C. L. Henley, Phys. Rev. B {\bf 69}, 075111 (2004). 
	\bibitem{BCMd} T. Barthel, M.-C. Chung and U. Schollwöck, Phys. Rev. A {\bf 74}, 022329 (2006).
	\bibitem{hatsugai06} S. Ryu and Y. Hatsugai, Phys. Rev. B {\bf 73}, 245115 (2006).
	\bibitem{chung16} M.-C. Chung {\it et al.}, Scientific Reports {\bf 6}, 29172 (2016).
	% quantum wave func
	\bibitem{Carleo17} G. Carleo and M. Troyer, Science {\bf 355}, 602 (2017).
	\bibitem{Gao17} X. Gao and L.-M. Duan, Nature Communications {\bf 8}, 662 (2017).
	\bibitem{Deng17a} D.-L. Deng, X. Li, and S. Das Sarma, Phys. Rev. X {\bf 7}, 021021 (2017).
	\bibitem{Deng17b} D.-L. Deng, X. Li, and S. Das Sarma, Phys. Rev. B {\bf 96}, 195145 (2017).
	\bibitem{Nomura17} Y. Nomura, Andrew S. Darmawan, Y. Yamaji, and M. Imada, Phys. Rev. B {\bf 96}, 205152 (2017).
	\bibitem{Kaubruegger18} R. Kaubruegger, L. Pastori, and J. C. Budich, Phys. Rev. B {\bf 97}, 195136 (2018).
	\bibitem{Glasser18} I. Glasser, N. Pancotti, M. August, I. D. Rodriguez, and J. I. Cirac, Phys. Rev. X {\bf 8}, 011006 (2018).
	\bibitem{Choo18} K. Choo, G. Carleo, N. Regnault, and T. Neupert, Phys. Rev. Lett. {\bf 121}, 167204 (2018).
	\bibitem{Melko19} G. Melko, G. Carleo, J. Carrasquilla,	and J I. Cirac, Nature Physics {\bf 15}, 887 (2019).
	\bibitem{Ohtsuki20} T. Ohtsuki and T. Mano, J. Phys. Soc. Jpn. {\bf 89}, 022001 (2020).
	% simulation
	\bibitem{Arsenault14} L.-F. Arsenault, A. Lopez-Bezanilla, O. A. von Lilienfeld, and A. J. Millis, Phys. Rev. B, {\bf 90}, 155136 (2014).
	\bibitem{Arsenault15} L.-F. Arsenault, O. A. von Lilienfeld, and A. J. Millis, arXiv:1506.08858 (2015). 
	\bibitem{Broecker17} P. Broecker, J. Carrasquilla, R. G. Melko, and S. Trebst, Scientific Reports {\bf 7}, 8823 (2017).
	\bibitem{Ryczko19} K. Ryczko, D. Strubbe, and I. Tamblyn, Phys. Rev. A {\bf 100}, 022512 (2019).
	\bibitem{Sellier19} J. M. Sellier, G. M. Caron, and J. Leygonie, Journal of Computational Physics, {\bf 387}, 154 (2019).
	\bibitem{Suwa19} H. Suwa, J. S. Smith, N. Lubbers, C. D. Batista, G.-W. Chern, K. Barros, Phys. Rev. B {\bf 99}, 161107 (2019).
	% phase transition
	\bibitem{Nieuwenburg17} Evert P. L. van Nieuwenburg, Y.-H. Liu, and S. D. Huber, Nature Physics {\bf 13}, 435 (2017).
	\bibitem{Carrasquilla17} J. Carrasquilla and R. G. Melko, Nature Physics {\bf 13}, 431 (2017).
	\bibitem{Ohtsuki16} T. Ohtsuki and T. Ohtsuki, J. Phys. Soc. Jpn. {\bf 85}, 123706 (2016).
	\bibitem{wang16} L. Wang, Phys. Rev. B {\bf 94}, 195105 (2016).
	\bibitem{Tanaka17} A. Tanaka and A. Tomiya, J. Phys. Soc. Jpn. 86, 063001 (2017).
	\bibitem{Wetzel17} S. J.Wetzel and M. Scherzer, Phys. Rev. B {\bf 96}, 184410 (2017).
	\bibitem{Hu17} W. Hu, R. R. P. Singh, and R. T. Scalettar, Phys. Rev. E {\bf 95}, 062122 (2017).
	\bibitem{Broecker17b} P. Broecker, F. F. Assaad, and S. Trebst, arXiv:1707.00663 (2017).
	\bibitem{Chng18} K. Ch’ng, N. Vazquez, and E. Khatami, Phys. Rev. E {\bf 97}, 013306 (2018).
	\bibitem{Liu18} Y.-H. Liu and E. P. L. van Nieuwenburg, Phys. Rev. Lett. {\bf 120}, 176401 (2018).
	\bibitem{Scheurer19} J. F. Rodriguez-Nieva and M. S. Scheurer, Nature Physics  {\bf 15}, 790–795 (2019).
       \bibitem{Scheurer20} M. Scheurer and R-J Slager, Phys. Rev. Lett. {\bf 124}, 226401 (2020).
	% topological phase transition
	\bibitem{kim17a} Y. Zhang and E.-A. Kim, Phys. Rev. Lett. {\bf 118}, 216401 (2017). 
	\bibitem{kim17b} Y. Zhang, R. G. Melko, and E.-A. Kim, Phys. Rev. B {\bf 96}, 245119 (2017).
	\bibitem{Zhang18} P. Zhang, H. Shen, and H. Zhai, Phys. Rev. Lett. {\bf 120}, 066401 (2018).
	\bibitem{Sun18} N. Sun, J. Yi, P. Zhang, H. Shen, and H. Zhai, Phys. Rev. B {\bf 98}, 085402 (2018).
	\bibitem{Carvalho18} D. Carvalho, N. A. Garca-Martnez, J. L. Lado, and	J. Fernandez-Rossier, Phys. Rev. B {\bf 97}, 115453 (2018).
	\bibitem{Ming19} Y. Ming, C.-T. Lin, S. D. Bartlett, and W.-W. Zhang, NPJ Computational	Materials {\bf 5}, 88 (2019).
	\bibitem{Caio19} M. D. Caio, M. Caccin, P. Baireuther, T. Hyart, and M. Fruchart, arXiv:1901.03346 (2019)
	\bibitem{Greplova20} E. Greplova, A. Valenti, G. Boschung, F. Schfer, N. Lrch, and S. D. Huber, New Journal of Physics {\bf 22}, 045003	(2020)..
	\bibitem{Zhang21} L.-F. Zhang, L.-Z. Tang, Z.-H. Huang, G.-Q. Zhang, W. Huang, D.-W. Zhang, Phys. Rev. A {\bf 103}, 012419 (2021).
    % explainable
    \bibitem{Zhang20} Y. Zhang, P. Ginsparg, and E.-A. Kim, Phys. Rev. Research {\bf 2}, 023283 (2020).
	\bibitem{tsai20} Y.-H. Tsai, M.-Z.  Yu, Y.-H. Hsu, and M.-C. Chung, Phys. Rev. B {\bf 102} 054512 (2020). 
	\bibitem{Dawid20} A. Dawid, P. Huembeli, M. Tomza, M. Lewenstein, and A. Dauphin, New J. Phys. {\bf 22}, 115001 (2020).
	\bibitem{Wooters98} W. K. Wooters, Phys. Rev. Lett. {\bf 80}, 2245  (1998).
	\bibitem{pytorch} A. Paszke, S. Gross, S. Chintala, G. Chanan, E. Yang, Z. DeVito, Z. Lin, A. Desmaison, L. Antiga, and A. Lerer, NIPS 2017.
	\bibitem{GBC} For example, see I. Goodfellow, Y. Bengio, and A. Courville, ``Deep Learning'', MIT Press (2016) and references therein.
	\bibitem{Kingma15} D. P. Kingma and J. Ba, {\it Proceedings of the 3rd International Conference on Learning Representations (ICLR)} (Cornell University, Ithaca NY, 2015).
	\bibitem{GradCAM} R. R. Selvaraju, M. Cogswell, A. Das, R. Vedantam, D. Parikh, and D. Batra, {\it International Conference on Computer Vision} (ICCV’17) (IEEE, Venice Italy, 2017), arXiv:1610.02391.
	
\end{thebibliography}
\end{document}